\documentclass[%
 aip,
 jmp,%
 amsmath,amssymb,
preprint,%
]{revtex4-1}

\usepackage{graphicx}
\usepackage{dcolumn}
\usepackage{bm}

\usepackage[utf8]{inputenc}
\usepackage[T1]{fontenc}
\usepackage{mathptmx}
\usepackage{etoolbox}

\usepackage{siunitx}
\usepackage{lipsum}
\usepackage{xcolor}
\usepackage[
    colorlinks=true,
    linkcolor=blue,
    citecolor=blue,
    urlcolor=blue
]{hyperref}

\DeclareSIUnit\torr{Torr}
\DeclareSIUnit\oersted{Oe}

\begin{document}

\preprint{AIP/123-QED}

\title[Disentangling Spin Pumping and Two-Magnon Scattering Contributions to Gilbert Damping in YIG/V Bilayers]{Disentangling Spin Pumping and Two-Magnon Scattering Contributions to Gilbert Damping in YIG/V Bilayers}

\author{S. ElKady}
\author{A. Tlais}
\author{H. Reslan}
\author{S. Isber}
\author{M. Haidar}
\email{mh280@aub.edu.lb}
\affiliation{Department of Physics, American University of Beirut, Riad El-Solh,
Beirut, 1107-2020, Lebanon }%



\date{\today}

\begin{abstract}
In this work, we investigate the magnetic damping and spin pumping response of YIG-based bilayers incorporating vanadium (V) as the normal metal layer via broadband ferromagnetic resonance (FMR) measurements as a function of YIG thickness. We show that the apparent enhancement of the Gilbert damping in YIG/V bilayers cannot be solely attributed to spin pumping. Instead, two-magnon scattering (TMS) plays a dominant role in governing the thickness dependence of the damping in the nanometer regime. By applying a thickness-dependent damping model that accounts for both spin pumping and two-magnon scattering contributions, we successfully disentangle the different relaxation contributions. Our analysis reveals that neglecting two-magnon scattering leads to an overestimation of the spin-pumping contribution and consequently to unphysically large values of the effective spin-mixing conductance. After isolating the intrinsic spin pumping contribution, we extract a thickness-independent effective spin-mixing conductance of $g^{\uparrow\downarrow}_{\mathrm{eff}} = 1.18 \times 10^{18}~\mathrm{m^{-2}}$. These findings provide a more accurate framework for quantifying spin transport parameters in ferromagnet/ heavy metal (FM/HM) systems and emphasize the necessity of accounting for extrinsic damping mechanisms when interpreting spin pumping and inverse spin Hall effect experiments.

\end{abstract}

\maketitle


\section{\label{sec:level1} Introduction}

\begin{figure*}[!ht]
\begin{center}
\includegraphics[width=0.8\textwidth]{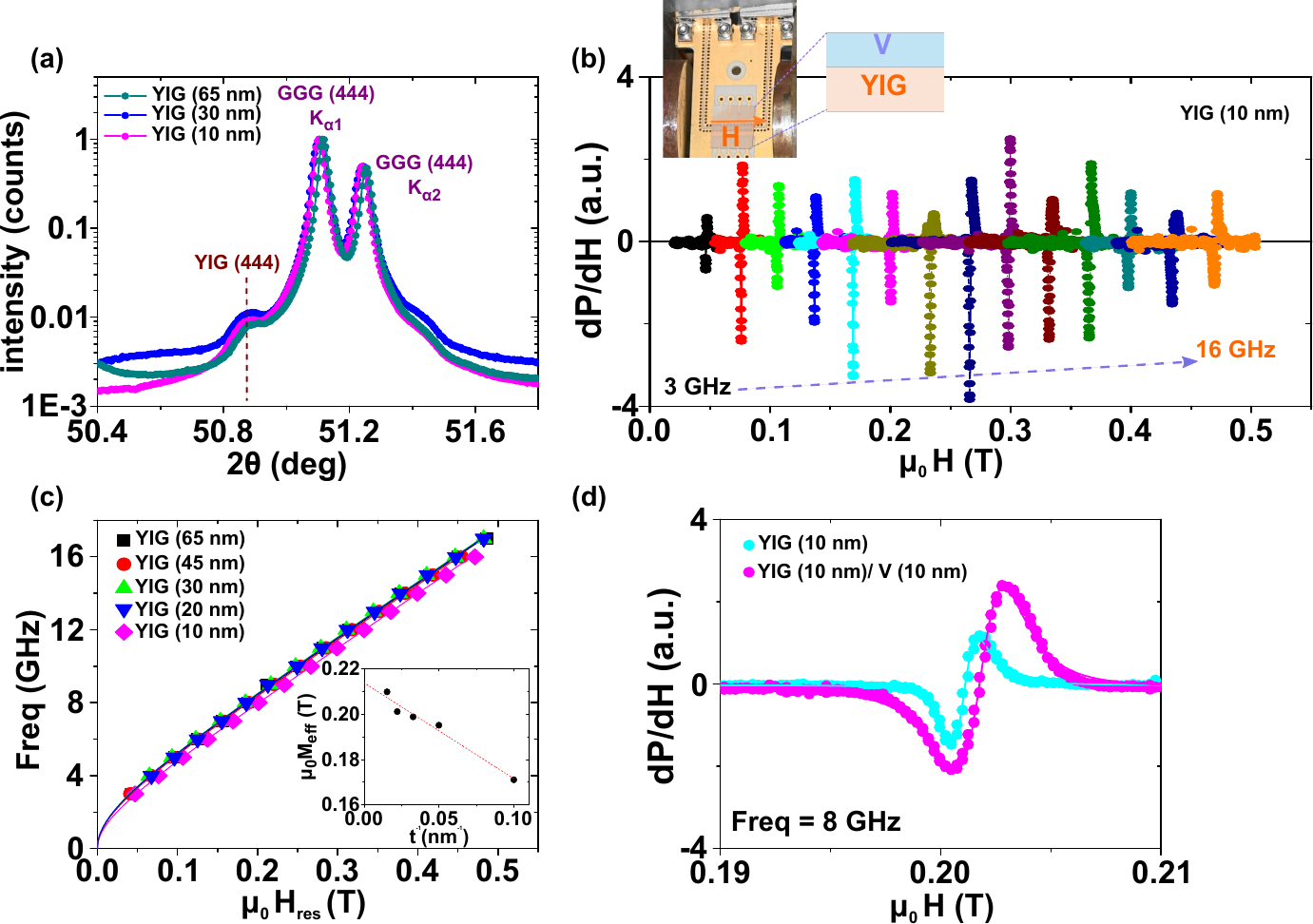}
\caption{(Color online) (a) X-ray diffraction (XRD) patterns of YIG films with different thicknesses, showing the characteristic YIG(444) reflection. (b) Schematic illustration of the broadband ferromagnetic resonance (B-FMR) measurement setup (inset), showing the coplanar waveguide, the in-plane applied magnetic field, and the YIG/V bilayer structure. Representative FMR spectra of a 10-nm-thick YIG film measured over the frequency range from \SI{3}{GHz} to \SI{16}{GHz} in \SI{1}{GHz} steps (main panel). 
(c) Frequency dependence of the resonance field for YIG films with different thicknesses. The solid lines correspond to fits using Kittel's equation. The inset shows the thickness dependence of the effective magnetization, $\mu_0 M_{\mathrm{eff}}$. (d) Comparison of the FMR spectra for a bare YIG film (cyan) and a YIG/V bilayer (magenta). The solid red curves represent fits using Eq.~(2).}.
\end{center}
\end{figure*}

Spintronics has emerged as a promising field for next-generation electronic technologies by exploiting both the charge and spin degrees of freedom of electrons \cite{Dieny2020}. Central to spintronic research is the understanding of spin current generation, manipulation, and detection, as well as the transfer of spin angular momentum across interfaces between ferromagnetic (FM) and normal metal (NM) materials\cite{Tserkovnyak2005, Zhang2015}. Among the key phenomena enabling spin-current-based functionalities are the spin Hall effect (SHE) \cite{D'yakonov1971,Hirsh1999} and its reciprocal process, the inverse spin Hall effect (ISHE). The SHE originates from spin–orbit coupling, which causes electrons with opposite spin orientations to deflect in opposite transverse directions, thereby converting a charge current into a pure spin current \cite{Liu2011,Hoffmann2013,Ranjbar2014,Chen2024,Haidar2025}. Conversely, the ISHE enables the conversion of spin currents into measurable charge currents, providing an efficient electrical probe of spin dynamics in magnetic systems \cite{Ando2008,AllivyKelly2013,Hahn2013_2,Balinsky2015,Haidar2016}. 
One of the most important mechanisms for generating spin currents is spin pumping in ferromagnetic resonance (FMR), which has been studied over the past two decades \cite{Yoshino2011, AllivyKelly2013,Caminale2016,Heinrich2011,Haertinger2015, Haidar2021, Mishra2025,Sahoo2025}. In an FM/NM bilayer, when the ferromagnetic layer is excited at resonance by a radio-frequency magnetic field under an external static magnetic field, its magnetization begins to precess. This dynamic precession transfers spin angular momentum across the FM/NM interface, generating a pure spin current in the adjacent normal metal \cite{Wang2013, Yang_2018,HARAJLI2024}. 
The generated dc spin current density can be written \cite{Mosendz2010, Weiler2013} as $J_{s}=\frac{\hbar \omega}{4\pi} Re(g^{\uparrow \downarrow}) P \sin^2 \theta,$ where $\omega =2\pi f$ is the angular precession frequency corresponding to the microwave frequency ($f$), $Re(g^{\uparrow \downarrow})$ represents the real part of the interfacial spin-mixing conductance, $\theta$ is the magnetization precession cone angle, and $P$ is a correction factor accounting for the ellipticity of the magnetization precession. The interfacial spin-mixing conductance, $g^{\uparrow \downarrow}$, has units of $m^{-2}$ and characterizes the efficiency of spin transmission across the FM/NM interface by describing the number of available spin transport channels per unit area \cite{Pai2015}. Theoretically, $g^{\uparrow \downarrow}$ can be calculated from the electronic band structure of FM/NM heterostructures \cite{Carva2007,Dolui2020}. Experimentally, spin pumping leads to an additional loss of angular momentum from the ferromagnet, thereby enhancing magnetic damping. Therefore, the increase in the Gilbert damping constant between the FM/NM bilayer and the bare FM layer can be used to determine the spin-mixing conductance through \cite{Sun2012,Jungfleisch2013}
\begin{equation}
g^{\uparrow \downarrow}=(4\pi M_{s} t_{FM})/(g \mu_{B} )(\alpha_{FM/NM}-\alpha_{FM} ),
\end{equation}
where $M_{s}$ is the saturation magnetization, $t_{FM}$ is the ferromagnetic layer thickness, $g$ is the Landé g-factor, and $\mu_{B}$ is the Bohr magneton. The Gilbert damping parameter is commonly extracted from the frequency dependence of the FMR linewidth. The quality of the FM/NM interface critically governs the efficiency of spin current generation and transfer. $g^{\uparrow \downarrow}$ is highly sensitive to interface properties such as cleanliness, defect density, structural quality, and magnetic uniformity. Since spin pumping originates from exchange interactions between the ferromagnetic moments and conduction electrons in the normal metal, the process is strongly dependent on atomic-scale interfacial characteristics \cite{Pogoryelov2020,Zhang2022, Jang2025,Bonda2026}. In ferromagnetic thin films at the nanometer scale, the magnetic damping is influenced not only by intrinsic Gilbert relaxation but also by additional interfacial mechanisms such as the spin memory loss (SML)\cite{RojasSanchez2014} and two-magnon scattering (TMS). This latter arises from the coupling of the uniform FMR mode to finite-wave-vector spin-wave modes due to magnetic inhomogeneities such as defects or surface roughness \cite{Hurben1998,Kalarickal2009}. These extrinsic processes contribute to an enhanced effective damping, which can mimic the thickness dependence associated with spin pumping and lead to an overestimation of spin transport parameters, particularly the spin-mixing conductance.

Although Pt is the most widely studied normal metal (NM) in YIG-based spin-pumping heterostructures, other heavy metals, including Ta \cite{Liu2011}, W \cite{Demasius2016}, WTi \cite{hachem2025}, and other 5d transition metals \cite{Wang}, have also attracted considerable interest. In particular, considerable efforts have been devoted to distinguishing the contributions of different mechanisms, including spin pumping and extrinsic damping, in Pt-based ferromagnetic bilayers \cite{Zhu2019, Zhu2020}; however, these mechanisms remain unexplored in YIG-based heterostructures. Vanadium (V) is a promising candidate because, despite its relatively weak spin–orbit coupling, it exhibits a substantial orbital Hall response \cite{Du2014_V_SD, LIU2024}. 

In this study, we investigate spin pumping in YIG/V bilayers using a 10-nm-thick V layer. We demonstrate that the enhancement of Gilbert damping in in-plane-magnetized YIG/V bilayers is not solely governed by spin pumping. Instead, TMS plays a dominant role in determining the dependence of the damping parameter $\alpha$ on the ferromagnetic layer thickness ($t_{FM}$). Neglecting TMS in the analysis of $\alpha$ can lead to unphysically large estimates of the spin-mixing conductance $g^{\uparrow\downarrow}$ and, consequently, inaccurate quantification of spin transport across the YIG/V interface. By properly accounting for the TMS contribution, we determine an effective spin-mixing conductance of $g^{\uparrow\downarrow}_{\mathrm{eff}} = 1.18 \times 10^{18}~\mathrm{m^{-2}}$ for the YIG/V bilayer.
More broadly, our results demonstrate that the separation of spin-pumping and TMS contributions is essential for reliably determining interfacial spin-transport parameters in YIG/normal-metal bilayers. By extending the role of TMS to a magnetic-insulator/normal-metal interface, our work highlights a general consideration for the interpretation of damping-based spin-pumping measurements in YIG-based devices.

\begin{figure*}[!ht]
\begin{center}
\includegraphics[width=0.7\textwidth]{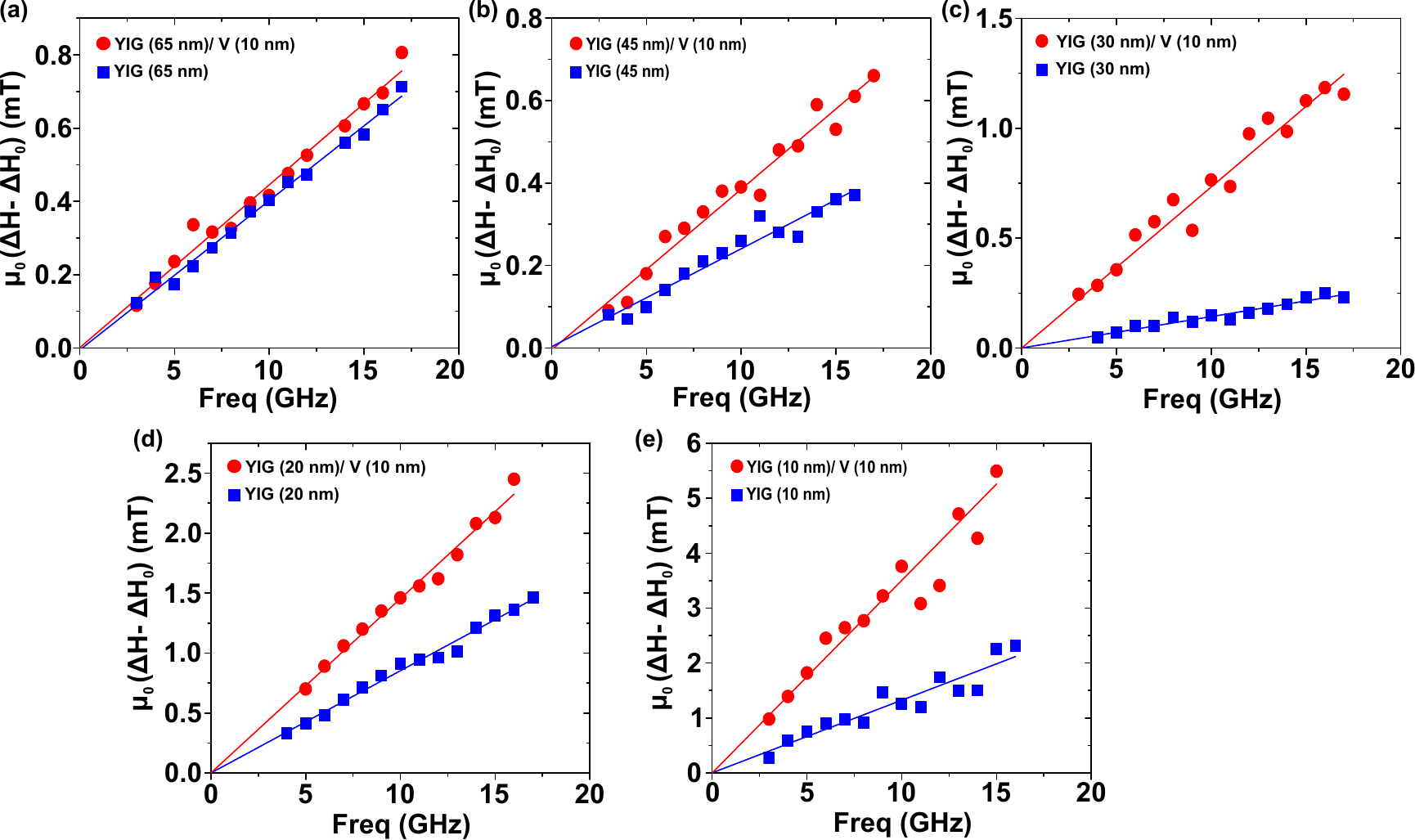}
\caption{(Color online) Frequency dependence of the FMR linewidth for bare YIG films (blue squares) and YIG/V bilayers (red circles) with different YIG thicknesses. The solid lines represent linear fits to the experimental data.}
\end{center}
\end{figure*}

\section{Experimental Setup}
For this study, yttrium iron garnet (YIG, $Y_{3}Fe_{5}O_{12}$) was selected as the ferromagnetic (FM) layer due to its exceptionally low magnetic damping and its ability to support pure spin-current transport without parasitic charge conduction \cite{Sun2012,Haidar2015}. The YIG thin films were grown on single-crystal (111)-oriented gadolinium gallium garnet (GGG, Gd3Ga5O12) substrates using pulsed laser deposition (PLD). A 248 nm KrF excimer laser operating at a pulse energy of 500 mJ and a repetition rate of 10 Hz was used to ablate the YIG target. The depositions were performed at a substrate temperature of 750,$^\circ\mathrm{C}$ under an oxygen pressure of 20 mTorr, while the base pressure of the chamber was maintained at $10^{-7}$ mbar. Additional details regarding the deposition process can be found in Ref. \cite{Haidar2023}. YIG films with thicknesses ranging from 10 to 65 nm (10, 20, 30, 45, and 65 nm) were deposited by controlling the deposition time. The film thicknesses were determined using a profilometer. 
Consequently, a 10-nm-thick V layer was deposited onto the YIG films by magnetron sputtering. This thickness falls within the reported spin-diffusion length range of V (5.2–16.3 nm), which depends on film and interface properties as well as the measurement technique \cite{Du2014_V_SD,Wang2017}. 
Fig. 1(a) presents the XRD patterns of three representative YIG films with thicknesses of 10, 30, and 65 nm. All exhibit the characteristic YIG (444) reflection near the GGG (444) substrate peak. Similar diffraction patterns were obtained for the other investigated thicknesses (20 and 45 nm), indicating that the YIG films maintain their crystalline quality and preferential epitaxial orientation on GGG(111) substrates throughout the thickness series.


Broadband ferromagnetic resonance (B-FMR) measurements were performed on 5×5 $mm^2$ samples using a coplanar waveguide to generate the microwave excitation field. The samples were mounted in a flip-chip configuration, with the external magnetic field applied in-plane, as illustrated in the inset of Fig.~1(b). The FMR response was detected using a lock-in amplifier while sweeping the magnetic field at fixed microwave frequencies ranging from 3 to 16 GHz with an input power of 0 dB. The measured spectra, recorded as the field derivative of the absorbed microwave power (dP/dH), exhibited the characteristic FMR lineshape shown in Fig.~1(b). The spectra were fitted using the derivative of a Lorentzian function containing both symmetric and antisymmetric contributions:

\begin{equation}
\begin{split}
  \frac{dP}{dH} &=A\frac{-8 \Delta H^2 \cdot (H - H_{\text{res}})}{(4(H - H_{\text{res}})^2 + \Delta H^2)^2} \\
  &+ B\frac{ \Delta H \cdot (\Delta H^2 - 4(H - H_{\text{res}})^2)}{(4(H - H_{\text{res}})^2 + \Delta H^2)^2} 
   + C \cdot H + D
\end{split}
\end{equation}

where $H_{res}$ and $\Delta H$ denote the resonance field and linewidth, respectively. The parameters A and B correspond to the amplitudes of the symmetric and antisymmetric components, while $C$ and $D$ account for the linear background slope and offset.
\begin{table*}[!htp]
\centering
\begin{tabular}{c|c|c|c|c|c|c|c}
\hline
\textbf{Sample} & $\gamma/2\pi $& $\mu_0 M_eff$ & $\alpha_{0}$ & $(\Delta H_{0})_{YIG}$ & $\alpha_{meas-IP}$ & $(\Delta H_{0})_{YIG/V}$ & $\alpha_{meas-OOP}$  \\
& (GHz/T) & (T) & ($\times 10^{-4}$) & (mT)     & ($\times 10^{-4}$) & (mT) & ($\times 10^{-4}$)\\
\hline
YIG (65 nm) / V (10 nm) & 29.45 & 0.21 $\pm$ 0.01  & 6.1 $\pm$ 0.2 & 0.5 $\pm$ 0.1  & 6.8 $\pm$ 0.3 & 1.02 $\pm$ 0.2 & 6.4 $\pm$ 0.1 \\
YIG (45 nm) / V (10 nm) & 29.5 & 0.20 $\pm$ 0.01 & 3.5 $\pm$ 0.2 & 1.1 $\pm$ 0.1   & 5.7 $\pm$ 0.3  & 1.1 $\pm$ 0.3 & 4.0 $\pm$ 0.2\\
YIG (30 nm) / V (10 nm) & 30 & 0.198 $\pm$ 0.01 & 2.1 $\pm$ 0.2 & 1.2 $\pm$ 0.1  &2  10.3 $\pm$ 0.3 & 1.06 $\pm$ 0.2 & 7.7 $\pm$ 0.3\\
YIG (20 nm) / V (10 nm) & 20 & 0.195 $\pm$ 0.01 & 12.7$\pm$ 0.3 & 1.4 $\pm$ 0.1  &2  21.6 $\pm$ 0.8 & 0.7 $\pm$ 0.1 & 16.7 $\pm$ 0.5\\
YIG (10 nm) / V (10 nm) & 29.2 & 0.17 $\pm$ 0.01 & 19.1 $\pm$ 0.6 & 1.1 $\pm$ 0.2   & 48.5 $\pm$ 0.7 &  0.98 $\pm$ 0.4 & 31.2 $\pm$ 0.3\\
\hline
\end{tabular}
\caption{Extracted values of the gyromagnetic ratio, $\gamma/2\pi$, effective magnetization, $\mu_{0}M_{\mathrm{eff}}$, intrinsic Gilbert damping, $\alpha_{\mathrm{0}}$, and inhomogeneous linewidth broadening, $(\Delta H_{0})_{\mathrm{YIG}}$, for the bare YIG films, together with the effective damping measured with the magnetic field applied in-plane, $\alpha_{\mathrm{meas-IP}}$, and out-of-plane, $\alpha_{\mathrm{meas-OOP}}$, and inhomogeneous broadening, $(\Delta H_{0})_{\mathrm{YIG/V}}$, for the YIG/V bilayers.}
\label{tab:3.3}
\end{table*}

\section{\label{sec:level2} Results and Discussion}

We evaluate the magnetization of YIG films by extracting the frequency dependence of the resonance field as presented in Fig. 1(c), using Kittel's relation $\textit{f}=\mu_{0}\gamma/2\pi \sqrt{(H)(H+M_{eff})},$ where $\gamma/2\pi$ is the gyromagnetic ratio and $M_{eff}$ is the effective magnetization. The results are tabulated in Table. I. The inset of Fig. 1(c) presents the dependence of $\mu_{0} M_{eff}$, on the inverse film thickness (1/t). A clear reduction in magnetization is observed for thinner films, with $\mu_{0} M_{eff}$ decreasing from approximately 0.21 T to 0.17 T as the YIG thickness is reduced from 65 to 10 nm. The data exhibit a linear dependence on $1/t$, which can be described by $\mu_{0} M_{eff} = 0.21 - 0.422 \times \frac{1}{t}$ where $t$ is given in nanometers. The intercept of the linear fit corresponds to the saturation magnetization, yielding $\mu_{0} M_{s} =$ 0.21 T. The slope is related to the interfacial magnetic anisotropy through the term $\frac{2K_{s}}{\mu_{0} M_{s} t}$. From the fit, the interfacial anisotropy constant was estimated to be $K_{s} = 4.4 \frac{mJ}{m^{2}}$. This surface anisotropy is attributed to interfacial effects arising from the YIG/Vanadium interface. 
Fig. 1(d) shows a comparison between the FMR spectra for YIG (10 nm) film with (magneta) and without (cyan) the vanadium layer. One can observe a broad peak in the bilayers, indicating enhanced damping relative to the YIG films alone, partially due to the spin pumping. 

The FMR linewidth contains contributions from different relaxation mechanisms. The first contribution arises from the intrinsic Gilbert damping ($\alpha_{0}$), a viscous relaxation process responsible for the dissipation of energy and angular momentum into the lattice. The second contribution originates from two-magnon scattering ($\alpha_{TMS}$), where the uniform FMR mode (k=0) couples to degenerate spin-wave modes with finite wave vectors. In the case of purely Gilbert-type viscous damping, the linewidth ($\mu_{0}\Delta H$) exhibits a strictly linear frequency dependence:

\begin{equation}
\mu_0\Delta H = (\mu_0\Delta H_0)_{YIG} + \left(\frac{2\alpha_{0}}{\gamma}\right)2\pi f,
\end{equation}
where $(\mu_0\Delta H_0)_{YIG}$ represents the inhomogeneous linewidth broadening. The intrinsic Gilbert damping constant of the bare YIG film, $\alpha_{0}$, was determined by analyzing the frequency dependence of the ferromagnetic resonance (FMR) linewidth, $\mu_{0}\Delta H$, as shown by the blue squares in Fig.~2(a--e). For all bare YIG samples, the linewidth exhibits an excellent linear dependence on frequency over the 2–16 GHz range, with no observable deviation from the expected Gilbert-damping behavior. Although TMS can contribute to the linewidth and may exhibit a nontrivial frequency dependence, no significant TMS contribution is resolved within our experimental uncertainty. Therefore, the damping extracted from the bare YIG films is predominantly Gilbert-like. By fitting the experimental data to Eq.~(3), we extracted $\alpha_{0}$ and $(\mu_0\Delta H_0)_{YIG}$ for all YIG films, as summarized in Table~I. It is worth noting that the bare YIG films exhibit an increase in damping, particularly in the thinner-films, which may be associated with surface and interfacial disorder, magnetic inhomogeneities \cite{Sun2012,Haidar2015}.  \newline

\begin{figure*}[!htp]
\begin{center}
\includegraphics[width=0.8\textwidth]{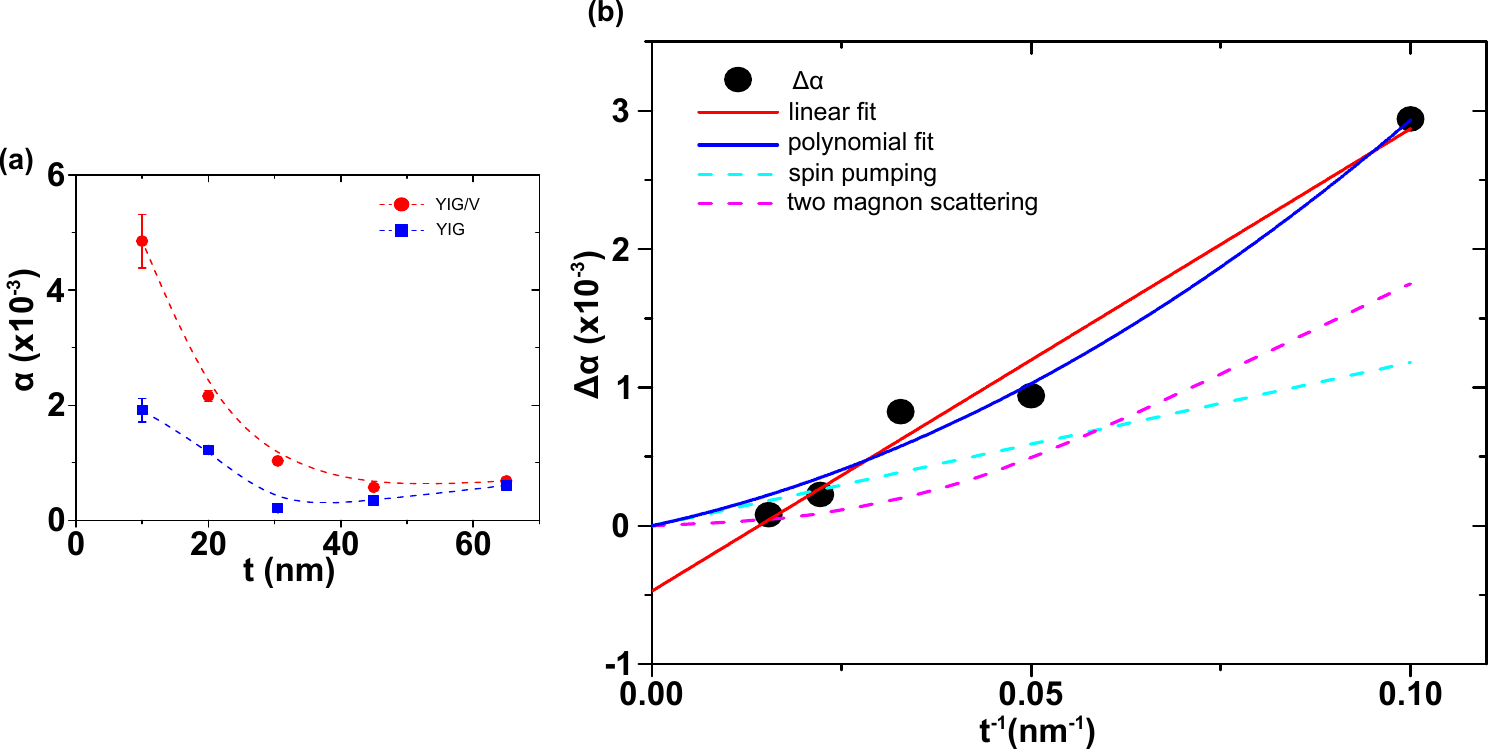}
\caption{(Color online)  (a) Thickness dependence of the extracted Gilbert damping parameter for bare YIG films and YIG/V bilayers. (b) Damping enhancement, $\Delta \alpha = \alpha_{\mathrm{meas}} - \alpha_{\mathrm{0}}$, plotted as a function of $t^{-1}$. The solid red and blue lines represent fits to the experimental data using Eqs.~(4) and (5), respectively. The dashed cyan and magenta curves separately show the contributions from spin pumping into the V layer and two-magnon scattering, illustrating their respective roles in the total damping enhancement. 
}
\end{center}
\end{figure*}

We then measured the FMR response of the YIG (t nm)/V (10 nm) bilayers and extracted the corresponding resonance linewidths for further determination of the effective damping. The red circles in Fig.~2(a--e) correspond to the frequency dependence of the linewidth measured for bilayers with variable YIG thickness. Compared to the bare YIG films, the bilayers exhibit a noticeably steeper $\Delta H/f$ slope, indicating an enhancement of the effective damping parameter, $\alpha_{\mathrm{meas}}$. The extracted values of $\alpha_{\mathrm{meas}}$ and $(\mu_0\Delta H_0)_{YIG/V}$ for the bilayer samples are also listed in Table~I.
Fig. ~3(a) presents the dependence of the damping parameter on the YIG film thickness for both bare YIG films ($\alpha_{0}$) and YIG/V bilayers ($\alpha_{meas}$). In both cases, the damping decreases with increasing YIG thickness. By analyzing the damping enhancement, $\Delta \alpha = \alpha_{meas} - \alpha_{0}$, it is evident that $\Delta \alpha$ increases significantly as the YIG thickness decreases. 
To first evaluate the spin pumping contribution, we initially assumed that the damping enhancement originates solely from spin pumping and modeled $\alpha_{sp}$ using a conventional inverse thickness dependence 
\begin{equation}
   \alpha_{meas} = \alpha_{0} + g^{\uparrow \downarrow}\frac{g \mu_{B}}{4\pi M_{s}}t^{-1}_{FM}. 
   \label{eq:sp}
\end{equation}  
where the first term corresponds to the intrinsic Gilbert damping, while the second term represents the additional damping contribution arising from spin pumping $\alpha_{sp}$. However, fitting the experimental data using Eq.~(4), shown by the solid red line in Fig.~3(b), resulted in an unphysical negative intercept. This indicates that the damping enhancement cannot be explained exclusively by spin pumping and that additional relaxation mechanisms must be considered. As a result, the spin-mixing conductance would be significantly overestimated.
TMS is known to contribute significantly to magnetic damping, particularly in thin films and at interfaces \cite{Azevedo2000}. Therefore in YIG/V bilayers, the effective damping enhancement could be modeled by considering contributions from both spin pumping and TMS as, 
\begin{equation}
\begin{aligned}
\alpha_{\mathrm{meas}} &= \alpha_{0} + \alpha_{\mathrm{sp}} + \alpha_{\mathrm{TMS}},\\
\alpha_{\mathrm{meas}} &= \alpha_{0} + g_{\mathrm{eff}}^{\uparrow\downarrow}
\frac{g\mu_{B}}{4\pi M_{s}}t_{\mathrm{FM}}^{-1}
+ \beta t_{\mathrm{FM}}^{-2}.
\end{aligned}
\label{eq:sp}
\end{equation}
The second term represents the interfacial contribution from spin pumping and spin memory loss, which scales as $1/t_{\mathrm{FM}}$ and is described by $g_{\mathrm{eff}}^{\uparrow\downarrow}$. The third term represents TMS contribution, which scales as $t_{\mathrm{FM}}^{-2}$ and is characterized by $\beta_{\mathrm{TMS}}$, which is related to $(K_s/M_s)^2$ and the density of magnetic defects at the FM surface. By fitting the experimental data with Eq.~(5), the spin-pumping and two-magnon scattering contributions could be quantitatively separated. The analysis reveals that the two-magnon scattering contribution, $\alpha_{\mathrm{TMS}}$, dominates the damping enhancement for thinner YIG films ($t < 30$ nm), while the spin pumping contribution, $\alpha_{\mathrm{sp}}$, remains comparatively smaller. In contrast, for thicker YIG films ($t > 30$ nm), the TMS contribution becomes significantly weaker and spin pumping becomes the dominant relaxation mechanism. The fitted dependence of the effective damping can be expressed as 
$\Delta \alpha = \alpha_{\mathrm{meas}} - \alpha_{0} = (0.0118 \pm 0.001~\mathrm{nm})\,t^{-1} + (0.175 \pm 0.02~\mathrm{nm}^{2})\,t^{-2},$ 
where the first term corresponds to spin pumping and the second term represents the two-magnon scattering contribution. The magnitude of the TMS contribution scales with $(\frac{2K_{s}}{M_{s}})^{2}$. Using the experimentally determined parameters for the YIG films, this quantity is estimated to be  $0.17 ~\mathrm{nm^ {2}}$, which is in good agreement with the model-derived coefficient of $0.175 ~\mathrm{nm^ {2}}$. To further illustrate the separation of these mechanisms, the extracted spin-pumping and two-magnon-scattering contributions are plotted individually in Fig.~3(b) as cyan dashed line and magenta curves, respectively. This analysis clearly demonstrates that neglecting two-magnon scattering leads to an overestimation of the spin-pumping contribution, particularly in ultrathin YIG films.

To independently assess the contribution of TMS in the YIG/V bilayers, we performed FMR measurements in the out-of-plane (OOP) configuration, where the TMS contribution is strongly suppressed. Fig.~4(a) shows the frequency dependence of $\Delta H-\Delta H_{0}$ obtained from the OOP measurements over the frequency range of 3--8~GHz. Fig.~4(b) presents the damping parameter as a function of YIG thickness for both the in-plane (IP, red closed circles) and OOP (magenta open symbols) configurations. The damping extracted from the OOP measurements is systematically lower than that obtained in the IP configuration, consistent with the suppression of TMS in the OOP measurements. In both configurations, the damping increases with decreasing YIG thickness, reflecting the increasing contribution of interfacial damping in thinner films. The TMS contribution is experimentally determined from the difference, $\alpha_{\mathrm{TMS}}=\alpha_{\mathrm{IP}}-\alpha_{\mathrm{OOP}}$ as shown in Fig~.5 (blue squares).

\begin{figure*}[!htp]
\begin{center}
\includegraphics[width=0.7\textwidth]{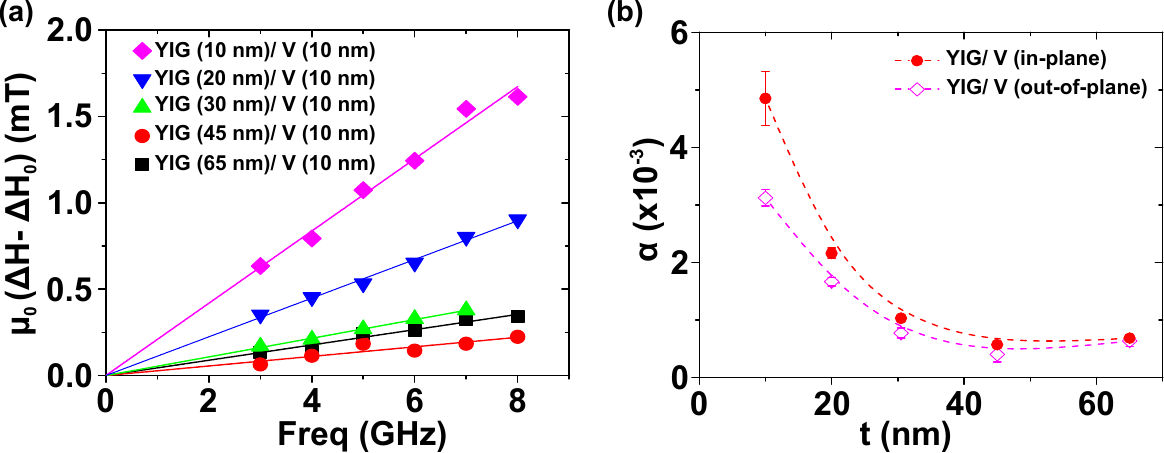}
\caption{(Color online) (a) Frequency dependence of $\Delta H-\Delta H_{0}$ obtained from the out-of-plane (OOP) FMR measurements of the YIG/V bilayers. (b) YIG-thickness dependence of the damping parameter measured in the in-plane (IP, red closed circles) and OOP (magenta open symbols) configurations. The dashedlines are guides to the eye.} 
\end{center}
\end{figure*}

To gain further insight into the origin of the TMS contribution and its dependence on the ferromagnetic layer thickness, we performed numerical calculations based on the Arias–Mills model \cite{Arias1999}. Within this framework, TMS originates from magnetic inhomogeneities at the ferromagnetic surface or interface, which can arise from surface defects, roughness, or local thickness variations. These perturbations break the spatial uniformity of the magnetic anisotropy and enable the uniform FMR mode to scatter into degenerate spin-wave modes with finite wave vectors. The model therefore provides a microscopic description of how interfacial inhomogeneities contribute to the measured FMR linewidth. The TMS contribution to the FMR linewidth broadening, $\Delta H_{\mathrm{TMS}}$, is expressed as
\begin{widetext}
\begin{equation}
\Delta H_{\mathrm{TMS}}
=
C
\left(
\frac{2K_s}{M_s}
\right)^2
\frac{1}{t_{\mathrm{FM}}^2}
\left[
\frac{
H_r^2
+
(H_r+4\pi M_{\mathrm{eff}})^2(a/c-1)
+
(4\pi M_{\mathrm{eff}})^2(c/a-1)
}{
(H_r+4\pi M_{\mathrm{eff}})^2
}
\right]
\sin^{-1}
\left(
\sqrt{
\frac{H_r}{H_r+4\pi M_{\mathrm{eff}}}
}
\right).
\end{equation}
\end{widetext}

where $H_r$ is the FMR resonance field and $4\pi M_{\mathrm{eff}} \approx 4\pi M_s+2K_s/(M_s t_{\mathrm{FM}})$ is the effective magnetization. The prefactor is given by $C=8b^2p/(\pi D)$, where $p$ represents the fraction of the surface covered by defects, $D$ is the exchange stiffness, and $a$, $b$, and $c$ characterize the dimensions of the surface defects. In our calculations, the experimentally determined magnetic parameters were used as inputs to the model. Specifically, we considered a YIG saturation magnetization of $M_s=1.67\times10^5~\mathrm{A/m}$, an interfacial anisotropy constant of $K_s=0.0044~\mathrm{J/m^2}$, extracted from our FMR measurements, and an exchange stiffness of $D=0.6\times10^{-12}~\mathrm{J,m^2}$. The remaining parameters describe the assumed geometry and distribution of the interfacial defects and were not directly extracted from the measurements. In particular, we used $b=0.3~\mathrm{nm}$ as a characteristic length scale for the local interfacial perturbations, $p=0.5$ as the assumed defect coverage, and $\langle a/c\rangle=\langle c/a\rangle=4$ to account for the finite lateral dimensions and aspect ratio of the defects. These values were adopted from from the original work of Arias and Mills as physically reasonable estimates to describe the interfacial inhomogeneities within the model.
The inset of Fig.~5 shows the calculated $\Delta H_{\mathrm{TMS}}$ as a function of frequency for different YIG thicknesses. The corresponding damping contribution due to TMS is extracted and plotted as a function of YIG thickness in the main panel of Fig.~5 (dashed line). Fig. 5 compares the TMS contributions as a function of YIG thickness determined using three approaches: (i) the difference between the in-plane (IP) and out-of-plane (OOP) FMR measurements (blue squares); (ii) the $t_{\mathrm{YIG}}^{-2}$ dependence extracted from the thickness-dependent damping analysis (red circles); and (iii) the Arias–Mills model (dashed line). The excellent agreement among these independent approaches provides compelling evidence for the significant contribution of TMS in YIG/V bilayers and supports its characteristic $t_{\mathrm{YIG}}^{-2}$ thickness dependence. These findings highlight the importance of accounting for TMS when extracting the spin-pumping contribution from FMR measurements in magnetic-insulator/normal-metal bilayers.

\begin{figure}[!htp]
\begin{center}
\includegraphics[width=0.4\textwidth]{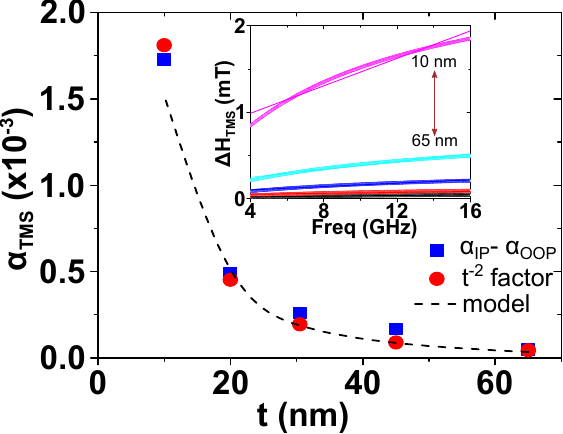}
\caption{(Color online) TMS contribution to the damping, $\alpha_{\mathrm{TMS}}$, as a function of YIG thickness, $t_{\mathrm{YIG}}$, extracted using three approaches: (i) the difference between in-plane (IP) and out-of-plane (OOP) FMR damping (blue squares), (ii) the $t_{\mathrm{YIG}}^{-2}$ dependence obtained from the thickness-dependent damping analysis (red circles), and (iii) the Arias–Mills model (black dashed lines). Inset: Calculated TMS-induced linewidth broadening versus frequency for different YIG thicknesses using the Arias–Mills model.}
\end{center}
\end{figure}

The spin memory loss (SML) mechanism is most significant in systems with strong interfacial spin–orbit coupling (ISOC), such as Co/Pt \cite{Zhu2019_SOT} and Co/AuPt \cite{Yun2024} bilayers, where ISOC reduces the spin transparency of the interface. Since SML originates from spin-flip processes associated with interfacial spin–orbit coupling, it is expected to be less pronounced at YIG/V interfaces than at interfaces involving heavy metals with stronger spin–orbit coupling, such as Pt \cite{RojasSanchez2014}.

\begin{figure}[!ht]
\begin{center}
\includegraphics[width=0.4\textwidth]{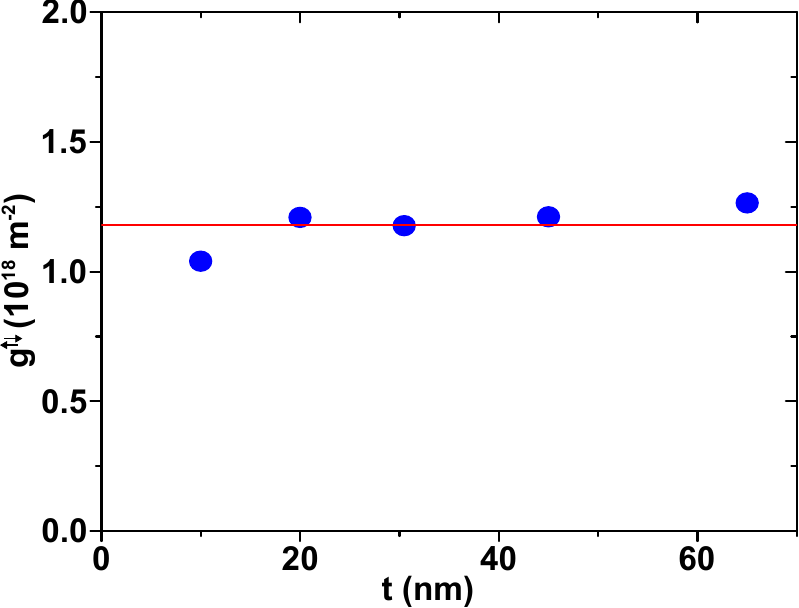}
\caption{(Color online) Dependence of the effective spin-mixing conductance on the YIG film thickness.}

\end{center}
\end{figure}

After isolating the spin pumping contribution to the damping enhancement in the YIG/V bilayers, the effective spin-mixing conductance was extracted for all film thicknesses, as shown in Fig.~6. The resulting values remain nearly constant over the entire thickness range, yielding an average $g^{\uparrow\downarrow}_{\mathrm{eff}}$ of $(1.18 \pm 0.05) \times 10^{18}~\mathrm{m^{-2}}$. This value is approximately three times lower than the previously reported value for YIG/V systems \cite{Du2015}. This discrepancy suggests that earlier spin-pumping analyses may have overestimated the spin-pumping contribution to the effective damping and, therefore, overestimated the spin diffusion lengths in heavy metals, which in turn led to an underestimation of their spin Hall angles in spin pumping and inverse spin Hall effect (ISHE) experiments.

In conclusion, our results demonstrate that the effective damping in YIG/V bilayers has a sizeable contribution from two-magnon scattering (TMS), rather than spin pumping, as the dominant mechanism governing the thickness dependence of the damping parameter. Neglecting the contributions of TMS, can therefore lead to unphysically large values of the effective spin-mixing conductance. Consequently, accurate determination of the spin Hall angle $\theta_{\mathrm{SH}}$ and spin Hall conductivity $\sigma_{\mathrm{SH}}$ requires a careful and quantitative separation of these damping mechanisms.

\begin{acknowledgments}
The authors would like to acknowledge the financial support of the American University of Beirut Research Board (URB), the IEEE Magnetics Society Hardship Countries Initiative, and the Research Funding for the Natural and Experimental Sciences from the Faculty of Arts and Sciences.
\end{acknowledgments}

\textbf{DATA AVAILABILITY}
The data that support the findings of this study are available from the corresponding author upon reasonable request.



%

\end{document}